\documentclass[12pt]{article}

\usepackage{amssymb,amsmath}
\usepackage{graphicx}

\newcommand{\be}{\begin{equation}}
\newcommand{\ee}{\end{equation}}

\newcommand{\dlt}{\delta}
\newcommand{\prt}{\partial}
\newcommand{\br}{{\bf r}}
\newcommand{\bk}{{\bf k}}

\newcommand{\ba}{{\bf a}}
\newcommand{\bu}{{\bf u}}
\newcommand{\bp}{{\bf p}}
\newcommand{\bbe}{{\bf e}}

\newcommand{\bt}{\beta}
\newcommand{\vp}{\varphi}

\newcommand{\al}{\alpha}
\newcommand{\ra}{\rightarrow}

\newcommand{\gm}{\gamma}
\newcommand{\om}{\omega}
\newcommand{\Om}{\Omega}

\newcommand{\dgr}{\dagger}

\newcommand{\Lbd}{\Lambda}

\newcommand{\cH}{{\cal H}}

\newcommand{\rgl}{\rangle}
\newcommand{\lgl}{\langle}

\begin{document}

\begin{center}

{\Large{\bf Double-Well Optical Lattices with Atomic Vibrations 
and Mesoscopic Disorder} \\ [5mm] 
 
V.I. Yukalov$^{1}$ and E.P. Yukalova$^{2}$} \\ [3mm]

{\it 
$^1$Bogolubov Laboratory of Theoretical Physics, \\
Joint Institute for Nuclear Research, Dubna 141980, Russia \\ [3mm]

$^2$Laboratory of Information Technologies, \\
Joint Institute for Nuclear Research, Dubna 141980, Russia} 

\end{center}

\vskip 2cm

\begin{abstract}

Double-well optical lattice in an insulating state is considered. The influence
of atomic vibrations and mesoscopic disorder on the properties of the lattice
are studied. Vibrations lead to the renormalization of atomic interactions. 
The occurrence of mesoscopic disorder results in the appearance of first-order
phase transitions between the states with different levels of atomic imbalance.
The existence of a nonuniform external potential, such as trapping potential,
essentially changes the lattice properties, suppressing the disorder fraction
and rising the transition temperature. 

\end{abstract}

\vskip 1cm

\newpage

\section{Introduction}

Optical lattices provide a convenient tool for modelling various many-body 
periodic structures [1-4]. The attractiveness of optical lattices is caused by 
the possibility of changing their properties in a required way. Thus, the lattice
period, the depth of potential wells at lattice cites, and the lattice geometry,
all these characteristics can be rather easily varied in experiment by applying
the necessary laser beams. Also, the strength of atomic interactions can be 
regulated in a wide range by the Feshbach resonance techniques [5,6]. Respectively, 
it is feasible to create many artificial periodic structures with the properties 
necessary for different applications. 

A special class of optical lattices is represented by the double-well lattices,
where each lattice cite is formed by a double-well potential [7-16]. Such lattices,
similarly to single-well lattices, can be in superfluid or insulating state. In 
addition, they can exhibit other states that cannot arise in the single-well 
counterparts. For example, they can exhibit the ordered states with nonzero mean
atomic imbalance and with phase transitions between the ordered state and
a disordered state with zero average atomic imbalance [7-16]. Another unusual 
property of the double-well lattices is the possibility of forming mixed states,
where mesoscopic regions of the ordered phase are intermixed with mesoscopic 
regions of the disordered phase [17]. In a bulk system, or in a sufficiently 
large trap, these regions are randomly distributed in space. While, for atoms 
in a tight trap, the region distribution is governed by the trap geometry. The
latter would be similar to the radial distribution of superfluid fraction in a 
trap with a periodic crystalline-type atomic configuration [18].

The mixed state of a double-well optical lattice, with mesoscopic regions of 
different phases, exhibits a variety of interesting properties, distinguishing
such a state from the purely ordered or completely disordered states. In our 
previous publication [17], we considered a rigid lattice with no external fields
acting on the system. The aim of the present paper is twofold: to generalize 
the consideration by including external fields, and also to analyze the role
of atomic vibrations in a double-well optical lattice. In Sec. 2, we derive
a model of an insulating optical lattice in order to explain the physical 
meaning of all its terms and for explicitly demonstrating the origin of the
term representing the influence of external potentials. In Sec. 3, atomic 
vibrations are introduced and their role in the renormalization of atomic 
interactions is analyzed. In Sec. 4, we study a double-well optical lattice
with mesoscopic disorder in the presence of an external field. Section 5 
contains conclusions and discussion.           

Throughout the paper, we use the system of units with $\hbar = 1$ and 
$k_B = 1$.

\section{Insulating Double-Well Optical Lattice}

Deriving the model of a double-well optical lattice, we start with the 
standard Hamiltonian 
\be
\label{1}
 \hat H = \int \psi^\dgr(\br) H_1(\br) \psi(\br) \; d\br +
\frac{1}{2} \int \psi^\dgr(\br) \psi^\dgr(\br') \Phi(\br-\br') 
\psi(\br') \psi(\br) \; d\br d\br' \;  ,
\ee
where $\psi({\bf r})$ is a field operator of atoms, $\Phi({\bf r})$ is a pair 
interaction potential, and in the single-atom Hamiltonian
\be
\label{2}
 H_1(\br) = - \; \frac{\nabla^2}{2m} + U(\br) \;  ,
\ee
the external potential is a sum
$$
 U(\br) = V_L(\br) + U_{ext}(\br) \;  ,
$$
where the first term is the lattice potential, periodic with the lattice 
spacing ${\bf a}$, while the second term is not periodic, representing an 
additional external field, e.g., corresponding to a trapping potential.
Atoms can be either bosons or fermions.
 
Keeping in mind an insulating lattice, the field operator can be expanded 
over an orthonormal basis of localized orbitals,
\be
\label{3}
\psi(\br) = \sum_{nj} c_{nj} \psi_n(\br-\br_j) \;    ,
\ee
where $n$ is a quantum multi-index and ${\bf r}_j$ is a vector of atomic
location. With this expansion, Hamiltonian (1) transforms to the expression
\be
\label{4}
\hat H =\sum_{ij} \sum_{mn} E_{ij}^{mn} c_{mi}^\dgr c_{nj} +
\frac{1}{2} \sum_{\{ j\} } \sum_{\{ n \} } \Phi_{j_1j_2j_3j_4}^{n_1n_2n_3n_4}
c_{n_1 j_1}^\dgr  c_{n_2 j_2}^\dgr c_{n_3 j_3} c_{n_4 j_4} \; ,
\ee
in which 
\be
\label{5}
E_{ij}^{mn} \equiv 
\int \psi_m^*(\br-\br_i) H_1(\br) \psi_n(\br-\br_j) \; d\br
\ee
and $\Phi_{j_1 j_2 j_3 j_4}^{n_1 n_2 n_3 n_4}$ are the matrix elements of 
the interaction potential. As localized orbitals, one can take maximally
localized Wannier functions designed for double-well lattices [19].

Considering the lattice with unity filling factor, it is necessary to
impose the no-double-occupancy constraint
\be
\label{6}
\sum_n c_{nj}^\dgr c_{nj} = 1 \; , \qquad c_{mj} c_{nj} =  0 \; .
\ee
And in the case of an insulating lattice, the no-hopping condition
\be
\label{7}
c_{mi}^\dgr c_{nj} = \dlt_{ij} c_{mj}^\dgr c_{nj}
\ee
is valid. Then Hamiltonian (4) becomes
\be
\label{8}
 \hat H = \sum_j \sum_{mn} E_{jj}^{mn} c_{mj}^\dgr c_{nj} +
\frac{1}{2} \sum_{i\neq j} \sum_{\{ n \} } V_{ij}^{n_1n_2n_3n_4}
c_{n_1 i}^\dgr c_{n_2 j}^\dgr c_{n_3 j} c_{n_4 i} \;  ,
\ee
in which
$$
V_{ij}^{n_1n_2n_3n_4} \equiv \Phi_{ijji}^{n_1n_2n_3n_4} \pm
\Phi_{ijij}^{n_1n_2n_3n_4} \;  ,
$$
the upper sign corresponding to Bose and the lower, to Fermi statistics.
In the sums over atomic locations, one has $i \neq j$. To simplify the 
notation, we shall omit in what follows this inequality, setting instead
the condition 
$$
 V_{jj}^{n_1 n_2 n_3 n_4} \equiv 0 \; .
$$
 
Each lattice site is formed by a double-well potential. The hopping between
different lattice sites is absent in an insulating lattice. But the 
tunneling between the wells of a double-well potential in the same site is,
generally, present. Therefore, to take into account the tunneling, it is 
necessary to consider minimally two energy levels. In what follows, we take
two such lowest levels with $n = 1,2$. The necessity of taking at least two 
quantum states distinguishes the case of the double-well lattice from that of 
a lattice with single-well sites. The usual situation is when the ground-state
function is symmetric with respect to spatial inversion, while the next-level
state is antisymmetric, so that
$$
 \psi_1(-\br) = \psi_1(\br) \; , \qquad
\psi_2(-\br) = - \psi_2(\br) \;  .
$$
Because of this property, the off-diagonal matrix elements, such as 
$V_{ij}^{1112}$ or $V_{ij}^{2221}$, are exactly zero. Even if the above 
symmetry property does not hold, such off-diagonal matrix elements are usually 
much smaller than the diagonal ones of the type $V_{ij}^{1111}, V_{ij}^{2222},
V_{ij}^{1221}$, and $V_{ij}^{1122}$. The nonzero matrix elements enter the 
following formulas through the combinations
$$
A_{ij} \equiv \frac{1}{4} \left ( V_{ij}^{1111} + V_{ij}^{2222} + 
2V_{ij}^{1221} \right ) \; , \qquad
B_{ij} \equiv \frac{1}{2} \left ( V_{ij}^{1111} + V_{ij}^{2222} - 
2V_{ij}^{1221} \right ) \; ,
$$
\be
\label{9}
C_{ij} \equiv \frac{1}{2} \left ( V_{ij}^{2222} - V_{ij}^{1111} 
\right ) \; , \qquad I_{ij} = - 2V_{ij}^{1122} \;   .
\ee
The latter are symmetric with respect to the change of the indices:
$$
A_{ij} = A_{ji} \; , \qquad B_{ij} = B_{ji} \; , \qquad
C_{ij} = C_{ji} \; , \qquad I_{ij} = I_{ji} \; .
$$

Introducing the notations
$$
K_{ij}^{mn} \equiv \int \psi_m^*(\br-\br_i) \left ( - \; \frac{\nabla^2}{2m} 
\right ) \psi_n(\br-\br_j) \; d\br \; ,
$$
\be
\label{10}
U_{ij}^{mn} \equiv 
\int \psi_m^*(\br-\br_i) U(\br) \psi_n(\br-\br_j) \; d\br  \; ,
\ee
and invoking the symmetry properties, we have
\be
\label{11}
 K_{jj}^{mn} = \dlt_{mn} K_{jj}^{nn} \;  ,
\ee
which gives
\be
\label{12}
 E_{jj}^{mn} = \dlt_{mn} K_{jj}^{nn} + U_{jj}^{mn} \;  .
\ee
The matrix elements of the kinetic energy can be represented as
\be
\label{13}
 \frac{\bp_j^2}{2m} \equiv \frac{1}{2} \left ( K_{jj}^{11} + 
K_{jj}^{22} \right ) \; .
\ee
Also, we define
\be
\label{14} 
E_0 \equiv \frac{1}{2} \sum_j \left ( E_{jj}^{11} + E_{jj}^{22} 
\right ) = \sum_j \frac{\bp_j^2}{2m} + U_0 N \;   ,
\ee
where
\be
\label{15}
 U_0 \equiv \frac{1}{2N} \sum_j \left ( U_{jj}^{11} + U_{jj}^{22} 
\right )  \; .
\ee
The quantity
\be
\label{16}
H_j \equiv - E_{jj}^{12} - E_{jj}^{21} = - U_{jj}^{12} - U_{jj}^{21}
\ee
plays the role of an external field acting on the atom in a $j$-th lattice 
site. And 
\be
\label{17}
\Om_j \equiv E_{jj}^{22} - E_{jj}^{11} + \sum_i C_{ij} 
\ee
has the meaning of the tunneling frequency characterizing atomic tunneling 
between the wells of a double well. Depending on the shape of the double-well 
potential, this frequency can be varied in a wide range [20].

Then we use the operator transformation
$$
c_{1j}^\dgr c_{1j} = \frac{1}{2} + S_j^x \; , \qquad
c_{2j}^\dgr c_{2j} = \frac{1}{2} - S_j^x \; ,
$$
\be
\label{18}
c_{1j}^\dgr c_{2j} = S_j^z - i S_j^y \; , \qquad 
c_{2j}^\dgr c_{1j} = S_j^z + i S_j^y \; .
\ee
The operators
$$
S_j^x = \frac{1}{2} \left ( c_{1j}^\dgr c_{1j} - 
c_{2j}^\dgr c_{2j} \right ) \; , \qquad
S_j^y = \frac{i}{2} \left ( c_{1j}^\dgr c_{2j} - 
c_{2j}^\dgr c_{1j} \right ) \; ,
$$
\be
\label{19}
S_j^z = \frac{1}{2} \left ( c_{1j}^\dgr c_{2j} + 
c_{2j}^\dgr c_{1j} \right )
\ee
satisfy spin algebra, independently from the statistics of atoms, whether 
Bose or Fermi. Because of their spin algebra, these operators are called 
pseudospin operators. With these notations, Hamiltonian (8) reduces to 
the form  
\be
\label{20}
\hat H = E_0 - \sum_j \left ( \Om_j S_j^x + H_j S_j^z \right )
+ \sum_{ij} \left ( \frac{1}{2} \; A_{ij} + B_{ij} S_i^x S_j^x -
I_{ij} S_i^z S_j^z \right ) \;   .
\ee

According to their properties, the operator $S_j^x$ describes atomic 
tunneling between the wells of a double-well potential in a $j$-th lattice
site. The operator $S_j^y$ corresponds to the Josephson current between 
the wells. And the operator $S_j^z$ characterizes atomic imbalance of 
these wells [14-17].

\section{Vibrating Double-Well Optical Lattice}

Atoms, forming the lattice, are, of course, not immovable, but can vibrate
around their lattice sites. In this section, we study the role of such 
vibrations.

Each atomic vector ${\bf r}_j$ can be represented as the sum
\be
\label{21}
\br_j = \ba_j + \bu_j
\ee
of the mean atomic location  
\be
\label{22}
\ba_j \equiv \lgl \br_j \rgl
\ee
and an atomic deviation from the mean location, such that 
\be
\label{23}
 \lgl \bu_j \rgl = 0 \;  .
\ee
The latter condition follows directly from definitions (21) and (22).

The interaction matrix elements (9) are functions of the atomic vectors, 
so that, for instance,
\be
\label{24}
A_{ij} \equiv A(\br_{ij}) \qquad 
(\br_{ij} \equiv \br_i - \br_j ) \; ,
\ee
and, similarly, all other matrix elements (9). Since we are considering
an insulating lattice, the atomic deviations are supposed to be small, 
because of which the interaction matrix elements can be expanded in 
powers of these deviations, limiting the expansions by the second-order
terms, e.g., as
\be
\label{25}
 A_{ij} \cong A(\ba_{ij}) + \sum_\al A_{ij}^\al u_{ij}^\al \; -\; 
\frac{1}{2} \sum_{\al\bt} A_{ij}^{\al\bt} u_{ij}^\al u_{ij}^\bt \;  ,
\ee
where the relative deviation is
\be
\label{26}
\bu_{ij} \equiv \bu_i - \bu_j
\ee
and the notations are used:
$$
A_{ij}^\al \equiv \frac{\prt A(\ba_{ij}) }{\prt a_i^\al} \qquad
(\ba_{ij} \equiv \ba_i - \ba_j ) \; ,
$$
$$
 A_{ij}^{\al\bt} \equiv 
\frac{\prt^2 A(\ba_{ij}) }{\prt a_i^\al\prt a_j^\bt} = - \;
\frac{\prt^2 A(\ba_{ij}) }{\prt a_i^\al\prt a_i^\bt} \;  .
$$

In order to guarantee the validity of condition (23), one has to introduce
the grand Hamiltonian
\be
\label{27}
 H = \hat H - \sum_j {\bf\Lbd}_j \cdot \bu_j \;  ,
\ee
with the Lagrange multipliers $\Lambda_j$ cancelling in the Hamiltonian the 
terms linear in $u_j^\al$. Then, expanding the matrix elements in powers of the
relative deviations (26), we come to the grand Hamiltonian
$$
H = U_0 N + \sum_j \left ( \frac{\bp_j^2}{2m} \; - \; \Om_j S_j^x - 
H_j S_j^z \right ) \; + \; 
\sum_{ij} \left \{ \frac{1}{2} \; A(\ba_{ij}) - \; 
\frac{1}{4} \sum_{\al\bt} A_{ij}^{\al\bt} u_{ij}^\al u_{ij}^\bt + 
\right.
$$
\be
\label{28}
+ \left.
\left [ B(\ba_{ij}) -\; 
\frac{1}{2} \sum_{\al\bt} B_{ij}^{\al\bt} u_{ij}^\al u_{ij}^\bt 
\right ] S_i^x S_j^x  - \left [ I(\ba_{ij}) -\; 
\frac{1}{2}  \sum_{\al\bt} I_{ij}^{\al\bt} u_{ij}^\al u_{ij}^\bt 
\right ] S_i^z S_j^z \right \} .
\ee

To make the problem treatable, one needs to decouple the vibrational and atomic 
degrees of freedom. We use the following decoupling:
\be
\label{29}
u_{ij}^\al u_{ij}^\bt S_i^\gm S_j^\gm =
\lgl  u_{ij}^\al u_{ij}^\bt \rgl S_i^\gm S_j^\gm +  
u_{ij}^\al u_{ij}^\bt \lgl S_i^\gm S_j^\gm \rgl -
\lgl u_{ij}^\al u_{ij}^\bt \rgl \lgl S_i^\gm S_j^\gm \rgl \; .
\ee

The existence of atomic vibrations leads to the renormalization of atomic 
interactions. In turn, atomic interactions renormalize the phonon matrix 
\be
\label{30}
 \Phi_{ij}^{\al\bt} \equiv A_{ij}^{\al\bt} + 
2B_{ij}^{\al\bt} \lgl S_i^x S_j^x \rgl
- 2 I_{ij}^{\al\bt} \lgl S_i^z S_j^z \rgl \;  .
\ee
The renormalized atomic interactions are
$$
\widetilde B_{ij} \equiv B(\ba_{ij}) + 
\sum_{\al\bt} B_{ij}^{\al\bt} \lgl u_i^\al u_j^\bt - u_j^\al u_j^\bt\rgl \; ,
$$
\be
\label{31}
\widetilde I_{ij} \equiv I(\ba_{ij}) + 
\sum_{\al\bt} I_{ij}^{\al\bt} \lgl u_i^\al u_j^\bt - u_j^\al u_j^\bt\rgl \;   .
\ee
Also, we introduce the notation
\be
\label{32}
 \widetilde E_0 \equiv \left ( \frac{A}{2} + U_0 \right ) N +
\sum_{ij} \sum_{\al\bt} \left ( I_{ij}^{\al\bt} \lgl S_i^z S_j^z \rgl -
 B_{ij}^{\al\bt} \lgl S_i^x S_j^x \rgl \right ) 
\lgl u_i^\al u_j^\bt - u_j^\al u_j^\bt\rgl \; ,
\ee
in which
\be
\label{33}
 A \equiv \frac{1}{N} \sum_{ij} A(\ba_{ij} ) \;  .
\ee
As far as $A$ is a constant, we have
\be
\label{34}
\sum_i A_{ij}^{\al\bt} = - \; \frac{\prt^2 A}{\prt a_j^\al \prt a_j^\bt} = 
0 \; , \qquad
 \sum_{ij} A_{ij}^{\al\bt} u_j^\al u_j^\bt =  
\sum_{i} A_{ij}^{\al\bt} \sum_j u_j^\al u_j^\bt = 0 \;.
\ee
In the phonon matrix (30), the main term is $A_{ij}^{\alpha \beta}$. 
Therefore one has
\be
\label{35}
\left | \sum_{ij} \Phi_{ij}^{\al\bt} u_j^\al u_j^\bt \right | \ll
\left | \sum_{ij} \Phi_{ij}^{\al\bt} u_i^\al u_j^\bt \right | \;  .
\ee

Thus, Hamiltonian (28) can be represented as the sum
\be
\label{36}
 H = \widetilde E_0 + H_{ph} + H_{at} \;  ,
\ee
in which the first term is given by Eq. (32), the second term is the phonon
Hamiltonian
\be
\label{37}
H_{ph} = \sum_j \frac{\bp_j^2}{2m} + 
\frac{1}{2} \sum_{ij} \sum_{\al\bt} \Phi_{ij}^{\al\bt} u_i^\al u_j^\bt \; ,
\ee
and the third term is the atomic Hamiltonian
\be
\label{38}
 H_{at} = -\sum_j \left ( \Om_j S_j^x + H_j S_j^z \right ) +
\sum_{ij} \left ( \widetilde B_{ij} S_i^x S_j^x -
\widetilde I_{ij} S_i^z S_j^z \right ) \;   .
\ee

The phonon spectrum $\omega_{ks}$ is defined by the eigenvalue problem
\be
\label{39}
 \frac{1}{m} \sum_{j\bt} 
\Phi_{ij}^{\al\bt} e^{i\bk\cdot\ba_{ij} } e_{ks}^\bt =
\om_{ks}^2 e_{ks}^\al \;  ,
\ee
where ${\bf e}_{ks}$ is a polarization vector, $k$ is quasi-momentum, and
$s$, polarization index. The phonon operators $b_{ks}$ are introduced 
through the transformation 
$$
\bp_j = -\; \frac{i}{\sqrt{2N} } \sum_{ks}\; \sqrt{m\om_{ks} }\; \bbe_{ks}
\left ( b_{ks} - b_{-ks}^\dgr \right ) e^{i\bk\cdot\ba_j} \; , 
$$
\be
\label{40}
\bu_j = \frac{1}{\sqrt{2N} } \sum_{ks} \; 
\frac{\bbe_{ks}}{\sqrt{m\om_{ks}} } \; 
\left ( b_{ks} + b_{-ks}^\dgr \right ) e^{i\bk\cdot\ba_j} \;   .
\ee
Using this, the phonon Hamiltonian (37) becomes diagonal,
\be
\label{41}
 H_{ph} = \sum_{ks} \om_{ks} 
\left ( b_{ks}^\dgr b_{ks} + \frac{1}{2} \right ) \;  .
\ee
This allows us to find the deviation correlation function
\be
\label{42}
 \lgl u_i^\al u_j^\bt \rgl = \frac{\dlt_{ij}}{2N} \; 
\sum_{ks}\; \frac{e_{ks}^\al e_{ks}^\bt}{m\om_{ks} } \; 
\coth\left ( \frac{\om_{ks}}{2T} \right ) \;  .
\ee
Since in the effective atomic interactions (31), we have $i \neq j$,   
they can be written as
\be
\label{43}
 \widetilde B_{ij} = B(\ba_{ij}) - 
\sum_{\al\bt} B_{ij}^{\al\bt} \lgl u_j^\al u_j^\bt \rgl \; , \qquad 
\widetilde I_{ij} = I(\ba_{ij}) - 
\sum_{\al\bt} I_{ij}^{\al\bt} \lgl u_j^\al u_j^\bt \rgl \;  .
\ee
And the nonoperator term (32) becomes
\be
\label{44}
  \widetilde E_0 = \left ( \frac{A}{2} + U_0 \right ) N +
\sum_{ij} \sum_{\al\bt} \left ( B_{ij}^{\al\bt} \lgl S_i^x S_j^x \rgl -
I_{ij}^{\al\bt} \lgl S_i^z S_j^z \rgl \right ) 
\lgl u_j^\al u_j^\bt \rgl \; .
\ee
 
The equation for the phonon spectrum (39) can be rewritten as
\be
\label{45}
 \om_{ks}^2 = \frac{1}{m} \; \sum_j \sum_{\al\bt} 
\Phi_{ij}^{\al\bt} e_{ks}^\al e_{ks}^\bt e^{i\bk\cdot\ba_{ij}} \;  .
\ee
For illustration, let us consider a lattice with cubic symmetry. Then
we can define the effective spectrum $\omega_k$ through the average
\be
\label{46}
 \om_k^2 \equiv \frac{1}{d} \sum_{s=1}^d \om_{ks}^2 \;  ,
\ee
where $d$ is space dimensionality. This yields
\be
\label{47}
 \om_k^2 = - \; \frac{1}{m} \; \sum_j D_{ij} e^{i\bk\cdot\ba_{ij}} \; ,
\ee
with the dynamical matrix
\be
\label{48}
 D_{ij} \equiv -\; \frac{1}{d} \; \sum_{\al=1}^d \Phi_{ij}^{\al\al} \; .
\ee
In the long-wave limit, using the property
$$
\sum_j D_{ij} \cong 0 \; ,
$$
we have 
\be
\label{49}
\om_k^2 \simeq \frac{1}{2m} \sum_j D_{ij} (\bk \cdot \ba_{ij} )^2 \qquad
(k\ra 0 ) \;   .
\ee
For a cubic lattice, the long-wave phonon spectrum reduces to 
\be
\label{50}
 \om_k \simeq ck \qquad (k \ra 0 ) \;  ,
\ee
with the sound velocity  
\be
\label{51}
  c = \sqrt{\frac{Da^2}{2m} } \; ,
\ee
and the isotropic dynamical matrix
\be
\label{52}
 D \equiv \frac{1}{a^2} \; \sum_j D_{ij} \left ( a_{ij}^\al\right )^2 \; .
\ee
Here $a$ is the lattice parameter of the cubic lattice.

It is important to emphasize that phonons can arise only in the presence
of long-range atomic interactions. For example, in the case of local 
interactions [4], we have 
$$
\Phi_{ij} \approx U \exp \left ( - q_0^2 a_{ij}^2 \right ) \; ,
$$
where
$$
a_{ij} \equiv | \ba_{ij} | \; , \qquad q_0 a \ll 1 \;   ,    
$$
and $U$ and $q_0$ are positive real parameters. This leads to
$$
 \Phi_{ij}^{\al\al} \approx 
2q_0^2 \Phi_{ij} \left [ 1 - 2q_0^2\left ( a_{ij}^\al \right )^2 \right ]\; .
$$
Then the dynamical matrix (48) is negative:
$$
 D_{ij} \approx - 2q_0^2 \Phi_{ij} \qquad (\Phi_{ij} > 0 ) \;  ,
$$
which makes it impossible to define a real phonon frequency. 

If $\Phi_{ij}$ is not Gaussian, but simply exponential, as
$$
\Phi_{ij} \approx U \exp(-{\bf q}_0 \cdot\ba_{ij} ) \qquad 
( U > 0 ) \;   ,
$$
then 
$$
\Phi_{ij}^{\al\al} \approx - \left ( q_0^\al \right )^2 \Phi_{ij} \;  ,
$$
and the dynamical matrix is positive:
$$
D_{ij} \approx \frac{1}{d} \; q_0^2 \Phi_{ij} \qquad 
( \Phi_{ij} > 0 ) \;   ,
$$
making well defined a real phonon frequency.  

When the interactions are of power law, as
$$
 \Phi_{ij} \approx \frac{U}{a_{ij}^n} \qquad (n > 0 ) \;  ,
$$
then
$$
\Phi_{ij}^{\al\al} \approx \frac{n\Phi_{ij}}{a_{ij}^2} \left [
1 - \; \frac{(n+2)(a_{ij}^\al)^2}{a_{ij}^2} \right ] \;  .
$$
The latter, for nearest neighbors, gives
$$
D_{ij} \approx \frac{n\Phi_{ij}}{a^2} 
\left ( \frac{n+2}{d}\;  - \; 1  \right ) \;  .
$$
Real phonon spectrum exists for a positive dynamical matrix $D_{ij}$. This 
means that the power law requires the condition
$$
 n > d - 2 \;  .
$$

If the lattice is sufficiently large, one can resort to the replacement of
the summation over momenta by the integration over the Brillouin zone 
$\mathcal{B}$, so that
$$
 \frac{1}{V} \; \sum_k \longrightarrow 
\int_{{\cal B}} \frac{d\bk}{(2\pi)^d } \;  .
$$
Then, from Eq. (43), we have 
\be
\label{53}
 \lgl u_j^\al u_j^\bt \rgl = \frac{\dlt_{\al\bt} }{2m\rho}
\int_{{\cal B}} \frac{1}{\om_k} \; \coth \left ( \frac{\om_k}{2T} 
\right ) \; \frac{d\bk}{(2\pi)^d} \;  .
\ee
This defines the mean square deviation 
\be
\label{54}
 r_0^2 \equiv \sum_{\al=1}^d \; \lgl u_j^\al u_j^\al \rgl =
\lgl u_j^\al u_j^\al \rgl d \;  .
\ee

In this way, the effective atomic interactions (43), for a cubic lattice,
become
\be
\label{55}
 \widetilde B_{ij} = B(\ba_{ij}) - \; \frac{r_0^2}{d} 
\sum_\al B_{ij}^{\al\al} \;  , \qquad
\widetilde I_{ij} = I(\ba_{ij}) - \; \frac{r_0^2}{d} 
\sum_\al I_{ij}^{\al\al} \;  ,
\ee
which shows how atomic vibrations change the interaction of atoms.

\section{Double-Well Lattice with Mesoscopic Disorder}

Atomic vibrations can be called {\it microscopic} fluctuations, since they 
are related to the fluctuations of separate atoms around their lattice sites. 
As is shown in the previous section, such fluctuations renormalize the 
strength of atomic interactions. 

There exists another type of fluctuations, whose occurrence can lead to a
much more noticeable change of system properties. These fluctuations
are termed {\it mesoscopic}, because each fluctuation involves many atoms 
that move coherently. The fluctuations are called mesoscopic, since their 
typical linear size is much larger than the mean interatomic distance, but 
smaller than the system size. These fluctuations represent strong 
fluctuations of order parameters, as a result of which the system can be 
treated as a fluctuating mixture of different thermodynamic phases, say,
an ordered and a disordered phase. Then the appearance of such mesoscopic
fluctuations of competing phases can be understood as the occurrence of
mesoscopic disorder. This type of fluctuations can lead to an essential 
change of system properties, as has been demonstrated for several kinds of 
matter of different physical nature, e.g., for ferromagnets [21,22], 
antiferromagnets [23,24], ferroelectrics [25], crystals [26], and other 
materials reviewed in Ref. [27]. A somewhat close behavior happens for the 
clustering nuclear matter [28,29].

In our previous paper [17], we have considered the influence of such 
mesoscopic disorder on double-well optical lattices, when there are no 
external fields. In the present section, we generalize the consideration to 
the case of a nonzero external field. The latter can be caused, e.g., by the 
presence of a trapping potential. We study a fluctuating mixture of two 
phases, an ordered phase, where there exists an average large atomic 
imbalance between the wells of double-well potentials, and a disordered phase, 
where such an imbalance is small. In the absence of external fields, the 
atomic imbalance of the disordered phase is exactly zero. 

After averaging over various configurations, produced by mesoscopic 
fluctuations, as described in Refs. [17,27], we have, instead of Eq. (20), 
the effective Hamiltonian
\be
\label{56}
 \widetilde H = \hat H_1 \bigoplus \hat H_2 \;  ,
\ee
in which each term 
\be
\label{57}
H_\nu = w_\nu E_0 - w_\nu \sum_j \left ( \Om_j S_j^x + H_j S_j^z \right )
+ w_\nu^2 \sum_{i\neq j} \left ( \frac{1}{2} \; A_{ij} + 
B_{ij} S_i^x S_j^x - I_{ij} S_i^z S_j^z \right ) 
\ee
corresponds to the related phase labelled by the index $\nu = 1,2$. Each 
Hamiltonian part (57) acts on a weighted Hilbert space $\mathcal{H}_\nu$. 
And the total Hamiltonian (56) is defined on the fiber space
\be
\label{58}
 \cH = \cH_1 \bigotimes \cH_2 \;  .
\ee
The geometric weights $w_\nu$ of the phases are defined as the minimizers 
of the thermodynamic potential 
\be
\label{59}
 F = - T \ln {\rm Tr} e^{-\bt\widetilde H} \;  ,
\ee
under the normalization condition
$$
 w_1 + w_2 = 1 \; , \qquad 0 \leq w_\nu \leq 1 \;  .
$$

The quantities of interest are the average values of the tunneling intensity
\be
\label{60}
 x_\nu \equiv \frac{2}{N} \sum_j \; \lgl S_j^y \rgl_\nu \;  ,
\ee
Josephson current
\be
\label{61}
 y_\nu \equiv \frac{2}{N} \sum_j \; \lgl S_j^y \rgl_\nu \;  ,
\ee
and the mean atomic imbalance
\be
\label{62}
 s_\nu \equiv \frac{2}{N} \sum_j \; \lgl S_j^z \rgl_\nu \;  .
\ee
Here the index $\nu$ in the statistical averages implies the averaging with
respect to Hamiltonian (57). The atomic imbalance (62) plays the role of an 
order parameter distinguishing an ordered and a disordered phases. This 
definition implies the constraint
\be
\label{63}
s_1 > s_2 \;   .
\ee
When external fields are switched off, the order parameter of the 
disordered phase has to be zero, which assumes the limiting condition
\be
\label{64}
 \lim_{ \{ H_j\ra 0 \} } s_2 = 0 \;  .
\ee

To calculate the above quantities (60), (61), and (62), we need to invoke 
some approximation for the Hamiltonian terms containing pairs of pseudospins.
For this purpose, we resort to the Kirkwood approximation [30], according 
to which for a pair of operators, acting on the space $\mathcal{H}_\nu$,
one has the decoupling 
\be
\label{65}
 S_i^\al S_j^\bt = g_{ij}^\nu \left (
\lgl S_i^\al \rgl_\nu S_j^\bt + S_i^\al \lgl S_j^\bt \rgl_\nu -
\lgl S_i^\al \rgl_\nu \lgl S_j^\bt \rgl_\nu \right ) \qquad
(i \neq j) \;  ,
\ee
in which $g_{ij}^\nu$ is a pair correlation function and the index $\nu$ at
the angle brackets means the statistical averaging with the related 
Hamiltonian (57). This decoupling differs from the mean-field approximation
by the presence of the pair correlation function.  

In the formulas below, we shall employ the notations for the average
interaction intensities
\be
\label{66}
B_\nu \equiv \frac{1}{N} \sum_{i\neq j} g_{ij}^\nu B_{ij} \; , \qquad
I_\nu \equiv \frac{1}{N} \sum_{i\neq j} g_{ij}^\nu I_{ij} \;   .
\ee
We assume that the trap is sufficiently wide, so that the quantities
\be
\label{67}
\Om_j = \Om \qquad H_j = H_0
\ee
can be treated as independent from the lattice indices. Also, we introduce
the notations
\be
\label{68}
\widetilde E_\nu \equiv w_\nu E_0 + w_\nu^2 \; \frac{N}{2} \left ( A - \; 
\frac{B_\nu}{2} \; x_\nu^2 + \frac{I_\nu}{2} \; s_\nu^2 \right )
\ee
and 
\be
\label{69}
\widetilde\Om_\nu \equiv w_\nu \Om - w_\nu^2 B_\nu x_\nu \; , \qquad
\widetilde I_\nu \equiv w_\nu H_0 + w_\nu^2 I_\nu s_\nu \;  .
\ee
Then Hamiltonian (57) reduces to the form
\be
\label{70}
 \hat H_\nu = \widetilde E_\nu - \sum_j \left (
 \widetilde\Om_\nu S_j^x + \widetilde I_\nu S_j^z \right ) \; .
\ee

For the thermodynamic potential (59), we find
\be
\label{71}
 F = F_1 + F_2 \; , \qquad F_\nu = \widetilde E_\nu
- NT \ln \left ( 2 \cosh \; \frac{J_\nu}{2T} \right ) \;  ,
\ee
where
\be
\label{72}
 J_\nu \equiv \sqrt{ \widetilde\Om_\nu^2 + \widetilde I^2_\nu } \;  .
\ee
The tunneling intensity (60) becomes
\be
\label{73}
 x_\nu = \frac{\widetilde\Om_\nu}{J_\nu} \; \tanh \left (
\frac{J_\nu}{2T} \right ) \;  .
\ee
The Josephson current (61) in an equilibrium system is zero,
\be
\label{74}
  y_\nu = 0 \; .
\ee
And the atomic imbalance (62) reads as
\be
\label{75}
 s_\nu = \frac{\widetilde I_\nu}{J_\nu} \; \tanh \left (
\frac{J_\nu}{2T} \right ) \;   .
\ee
  
In order to estimate the relation between the different interaction terms (9),
we can take into account that the values $V_{ij}^{1111}, V_{ij}^{2222}$ and
$V_{ij}^{1221}$ are close to each other, so that we can set
$$
V_{ij}^{1111} =  V_{ij}^{2222} = V_{ij}^{1221} \equiv  V_{ij}\;  .
$$
As a result,
$$
A_{ij} = V_{ij} \; , \qquad B_{ij} = 0 \; , \qquad C_{ij} =0 \; .
$$
We shall use this simplification in what follows.

To accomplish numerical calculations, we need to define the pair correlation 
function. Here, for simplicity, we set
\be
\label{76}
 g_{ij}^\nu = 1 \;  ,
\ee
which reduces the Kirkwood approximation to the mean-field form. For the 
convenience of numerical calculations, we define the dimensionless thermodynamic 
potentials
\be
\label{77}
  G \equiv \frac{F}{NI} \; , \qquad G_\nu \equiv \frac{F_\nu}{NI} \; ,
\ee
in which
\be
\label{78}
 I \equiv \frac{1}{N} \sum_{i\neq j} I_{ij} = I_\nu \; .
\ee
It is also convenient to introduce the dimensionless parameters
\be
\label{79}
 u \equiv \frac{A}{I} \; , \qquad \om \equiv \frac{\Om}{I} \; , \qquad
h \equiv \frac{H_0}{I} \;  .
\ee
Then for quantities (69) and (72), we have
\be
\label{80}
 \frac{ \widetilde \Om_\nu}{I} = w_\nu \om \; , \qquad 
\frac{J_\nu}{I} = w_\nu f_\nu \; , \qquad 
\frac{\widetilde I_\nu}{I} = w_\nu h + w_\nu^2 s_\nu \; ,
\ee
where
\be
\label{81}
 f_\nu \equiv \sqrt{(h+w_\nu s_\nu)^2 + \om^2 } \;  .
\ee
 
Measuring temperature in units of $I$, for the tunneling intensity (73), 
we get
\be
\label{82}
 x_\nu = \frac{\om}{f_\nu}\; \tanh \left ( 
\frac{w_\nu f_\nu}{2T} \right ) \;  ,
\ee
while the atomic imbalance (75) becomes
\be
\label{83}
 s_\nu = \frac{h+w_\nu s_\nu}{f_\nu}\; \tanh\left (
\frac{w_\nu f_\nu}{2T} \right ) \;  .
\ee
  
In the dimensionless notation, the thermodynamic potential (77) reads as
\be
\label{84}
 G = G_1 + G_2 \; , \qquad
G_\nu = e_\nu - T \ln \left [ 2 \cosh\left ( 
\frac{w_\nu f_\nu}{2T} \right ) \right ] \;  ,
\ee
with
\be
\label{85}
e_\nu = w_\nu \; \frac{E_0}{NI} + 
\frac{w_\nu^2}{4} \left ( 2u + s_\nu^2 \right ) \;   .
\ee
 
Recall that, in addition to the above equations, we have to define the 
geometric weights of the ordered and disordered phases as the minimizers
of the thermodynamic potential $G$. To take explicitly into account the 
normalization condition, we introduce the notation
\be
\label{86}
w \equiv w_1 \; ,  \qquad w_2 = 1 - w \;  .
\ee
Minimizing the thermodynamic potential (84) with respect to $w$, we find 
the equation
\be
\label{87}
w = \frac{2u+\om(x_1-x_2) + h(s_1 - s_2) - s_2^2}{4u - s_1^2 -s_2^2} \;  ,
\ee
defining the weight of the ordered phase. The weight of the disordered phase,
according to relation (86), is $1-w$.

When the gap between the energy levels $E_{jj}^{11}$ and $E_{jj}^{22}$, 
defined in Eq. (5), is small, as compared to the interaction intensity (78), 
then, in view of definition (17), one has $\omega \ll 1$. In that case, 
we have
$$
f_\nu = h + w_\nu s_\nu \; , \qquad x_\nu = 0 \qquad ( \om\ra 0 ) \; ,
$$
$$
s_\nu = \tanh \left [ \frac{w_\nu(h + w_\nu s_\nu)}{2T} \right ] \; .
$$
Therefore, the atomic imbalance for the ordered phase becomes
\be
\label{88}
 s_1 = \tanh \left ( \frac{wh+ w^2 s_1}{2T} \right ) \;  .
\ee
The atomic imbalance for the disordered phase is
\be
\label{89}
s_2 = \tanh \left (\frac{(1- w)h+ (1-w)^2 s_2}{2T} \right ) \;  .
\ee
And the equation (87) for the geometric weight of the ordered phase takes 
the form
\be
\label{90}
 w = \frac{2u + h(s_1-s_2) - s_2^2}{4u - s_1^2 - s_2 ^2 } \;  .
\ee
These are the main equations we need to solve, keeping in mind constraints 
(63) and (64). In addition, we have to compare the thermodynamic potential 
of the mixed system, (84), with that of a pure ordered system, when $w = 1$
and there exists just one order parameter $s_1$ given by the equation
$$
s_1 = \tanh \left ( \frac{s_1+h}{2T} \right ) \;   .
$$
Note that there always exists a solution with $w = 0.5$ representing a 
degenerate case, when the competing phases are asymptotically equivalent.
Among all possible solutions, we have to select the most stable one,
which minimizes the system thermodynamic potential and satisfies all imposed
constraints, such as Eqs. (63) and (64) and the condition $0 \leq w \leq 1$. 
The solution $w = 0$ makes the value $s_1$ undefined, hence constraint (63)
inapplicable. Therefore this case is unphysical and has to be excluded.

The results of numerical calculations are presented in Figs. 1 and 2. The 
parameters $u$ and $h$ have been varied in a wide range. When $u \leq 0$, 
the most stable solution corresponds to the pure system, with no mesoscopic 
disorder and $w = 1$ for all temperatures. Positive values of $u$ allow for 
the appearance of mesoscopic disorder. According to definition (79), the 
positive parameter $u$ can be interpreted as the ratio of the disordering
interaction strength $A$ to the ordering interaction strength $I$. Thence
this ratio $u$ can be called the disordering parameter. It turns out that
for $u > 0.5$ the solutions are not continuous. This could mean that such 
a strong disorder parameter prohibits the states with mesoscopic fluctuations,
at least in the frame of the present model. To consider the higher values 
of $u$ requires to invoke a more sophisticated approximation than the 
mean-field one. Varying $u$ in the interval $0 < u < 0.5$ shows that at 
$T = 0$, we have $s_1 = 1$ and the most stable is the state with $w = 1$. 
With rising temperature, the state with $w=1$ and $s_2=0$ persists till 
the transition temperature $T_0$, where there occurs a first-order phase 
transition to the state with $w=0.5$. The transition temperatures, as 
functions of $u$ and $h$, are depicted in Fig 1. The behavior of the order 
parameter $s_1$ is presented in Fig 2 for different values of $u$ and $h$.

\section{Discussion}

The influence of atomic vibrations and mesoscopic disorder on the properties 
of an insulating double-well optical lattice are studied. Atomic vibrations 
result in the appearance of well-defined phonon degrees of freedom, provided
that the interactions between atoms are sufficiently long-ranged. For example,
local interactions, described by a delta-function potential, typical of many 
rarified trapped gases, do not allow for the formation of well-defined phonons,
since their spectrum turns out to be imaginary. But long-range interactions, 
such as existing between atoms with electric or magnetic dipoles, perfectly
allow for the appearance of well-defined phonon degrees of freedom. At the 
present time, several atomic species are known, enjoying such long-range dipole 
interactions, for instance, $^{52}$Cr possessing large magnetic moments [31,32], 
Rydberg atoms [33] and polar molecules [34].

So, the consideration of the present paper concerns this type of dipolar gases 
loaded into an optical lattice. We show that atomic vibrations in a double-well 
optical lattice renormalize the strength of atomic interactions.

Atomic vibrations can be called {\it microscopic} fluctuations, since they 
correspond to the oscillations of separate atoms, though interacting with each 
other, but the motion of atoms being not correlated. There exists another kind of 
fluctuations, when atoms move in a coherent manner, locally defining an order 
parameter that differs from that of the surrounding matter. The typical size 
of such a coherent atomic group is intermediate between the mean interatomic 
distance and the linear system size, because of which these coherent oscillations 
are termed {\it mesoscopic}. When the latter correspond to the order parameters
of different thermodynamic phases, these oscillations can be called 
{\it heterophase}. Such mesoscopic fluctuations can occur in various statistical
systems of condensed-matter type [21-27] and in many macromolecular systems [35].  

Mesoscopic fluctuations can also occur in double-well optical lattices [17]. 
Here, an ordered state corresponds to a nonzero atomic imbalance between the 
wells of the double wells. If there are no external fields, the disordered state
is characterized by zero atomic imbalance [17]. But in the presence of external 
fields, disorder corresponds to an atomic imbalance that is not exactly zero, 
but smaller than the imbalance in the ordered state. We have generalized the 
previous analysis of mesoscopic disorder in a lattice without external fields
to the case of nonzero external fields that can be related, e.g., to the 
presence of a trapping potential. The existence of external fields essentially
changes the system properties, leading to the occurrence of first-order phase 
transitions between the states with different order parameters, that is, 
between the states with different atomic imbalance.

It is worth noting that term (16), corresponding to an external field, arises
only when, in addition to the lattice potential, there is a nonuniform 
external potential, such as trapping potential. To show this, let us consider
an ideal optical lattice with a periodic potential
$$
U(\br) = U(\br + \ba_j) \;   .
$$
In that case, the single-atom Hamiltonian (2) enjoys Bloch waves as its 
eigenstates, defined by the eigenproblem
$$
H_1(\br) \vp_{nk}(\br) = E_{nk} \vp_{nk}(\br) \;   .
$$
Taking for atomic orbitals the well-localized Wannier functions [19], we have 
the relation
$$
\psi_n(\br-\ba_j) = 
\frac{1}{\sqrt{N}} \; \sum_k \vp_{nk}(\br) e^{-i\bk\cdot\ba_j} \; .
$$
Then for the matrix elements (5), we obtain
$$
E_{ij}^{mn} = \frac{\dlt_{mn}}{N} \sum_k E_{nk}  e^{i\bk \cdot\ba_j} \; .
$$
Thus, these matrix elements are purely diagonal with respect to the band 
indices $m,n$. Therefore, the effective field (16), caused by nondiagonal 
elements, is strictly zero, hence $h = 0$. 
  
The occurrence of mesoscopic fluctuations, causing the appearance of 
mesoscopic disorder in a double-well optical lattice, changes the system 
properties in a much more dramatic way than microscopic fluctuations, such 
as atomic vibrations.

\vskip 5mm

{\bf Acknowledgement}

\vskip 2mm
We are grateful for many useful discussions to V.S. Bagnato. Financial support 
from the Russian Foundation for Basic Research is appreciated.

\newpage

\begin{center}
{\Large{\bf Figure Captions}}
\end{center}

\vskip 1cm
Fig. 1. Transition temperature $T_0$ in units of $I$: (a) as a function of $h$ 
for different values of $u$; (b) as a function of $u$ for different values of $h$.

\vskip 1cm
Fig. 2. Order parameter $s_1$ as a function of dimensionless temperature for 
different values of the disorder parameter: (a) $u = 0.1$; (b) $u = 0.5$. The
lines are marked by the numbers corresponding to different values of the external 
field: (1) $h = 0$; (2) $h = 0.2$; (3) $h = 0.4$; (4) $h = 0.6$; (5) $h = 0.8$; 
(6) $h = 1.0$.  

\newpage

\begin{figure}[ht]
\vspace{9pt}
\centerline{
\hbox{ \includegraphics[width=8.5cm]{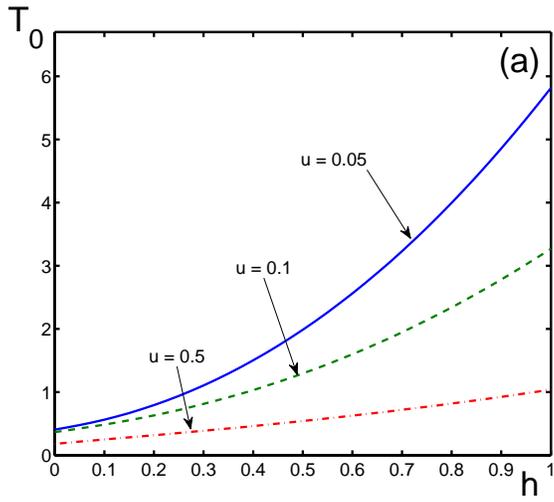} \hspace{1cm}
\includegraphics[width=8.5cm]{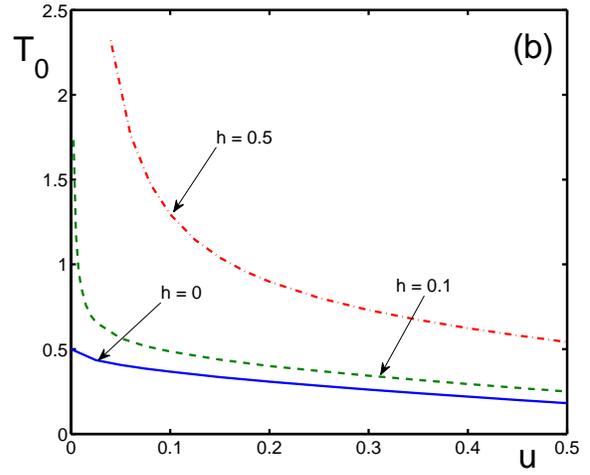} } }
\caption{Transition temperature $T_0$ in units of $I$: (a) as a function of $h$ 
for different values of $u$; (b) as a function of $u$ for different values of $h$.}
\label{fig:Fig.1}
\end{figure}

\begin{figure}[ht]
\vspace{9pt}
\centerline{
\hbox{ \includegraphics[width=8.5cm]{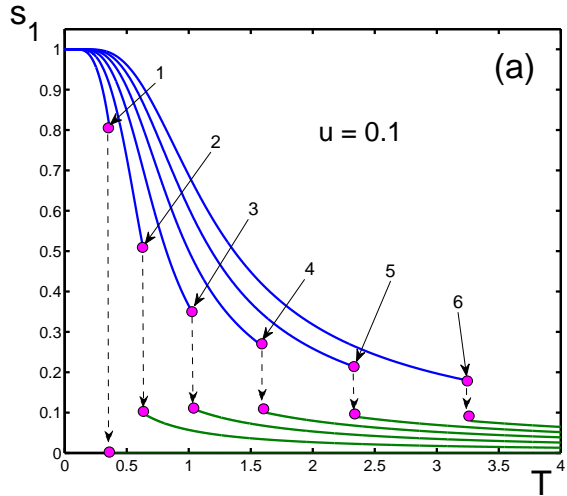} \hspace{1cm}
\includegraphics[width=8.5cm]{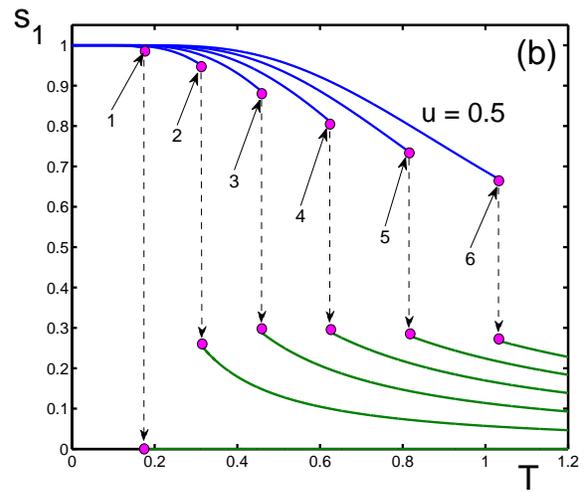} } }
\caption{Order parameter $s_1$ as a function of dimensionless temperature for 
different values of the disorder parameter: (a) $u = 0.1$; (b) $u = 0.5$. The
lines are marked by the numbers corresponding to different values of the external 
field: (1) $h = 0$; (2) $h = 0.2$; (3) $h = 0.4$; (4) $h = 0.6$; (5) $h = 0.8$; 
(6) $h = 1.0$.}
\label{fig:Fig.2}
\end{figure}
\end{document}